\documentclass[12pt]{article}
\DeclareMathSizes{12}{12}{10}{10}

\usepackage[margin=1in]{geometry}

\usepackage{amsmath, mathtools, graphicx}

\usepackage{mathrsfs, appendix}
\usepackage{nicefrac}
\usepackage{xfrac}

\newcommand\numberthis{\addtocounter{equation}{1}\tag{\theequation}}

\begin{document}


\begin{titlepage}
	\thispagestyle{plain}
	\begin{center}
    	\large
		\vspace*{0.6cm}
    	\textbf{Direct Interaction Approximation for Non-Markovianized Stochastic Models in the Turbulence Problem}

		\normalsize
    	\vspace{0.9cm}
    	B.K. Shivamoggi and N. Tuovila

		University of Central Florida

		Orlando, FL 32816-1364
\end{center}
    	\vspace{1.0cm}
    	\textbf{Abstract}
		\vspace{0.9cm}

		The purpose of this paper is to explore mathematical aspects associated with the application of the \textit{direct interaction approximation} (DIA) (Kraichnan \cite{Kraichnan-1961},\cite{Kraichnan-1965}) to the non-Markoviani-zed stochastic models in the turbulence problem.  This process is shown to lead to a functional equation, and construction of solutions of this equation is addressed within the framework of a continued fraction representation.  The relation of the DIA solution to the perturbative solution is discussed.  The DIA procedure is applied to the problem of wave propagation in a random medium, which is described by a stochastic differential equation, with the characteristics of the medium represented by stochastic coefficients.  The results are compared with those given by the perturbative procedure.

\end{titlepage}

\setcounter{page}{2}

\subsection*{1. Introduction}
The direct-interaction approximation (DIA) developed by Kraichnan \cite{Kraichnan-1961}, \cite{Kraichnan-1965} is currently the only fully self-consistent analytical theory of turbulence in fluids (Kraichnan \cite{Kraichnan-1965})\footnotemark.  One feature of the DIA is its physical realizability, which avoids the catastrophic behavior associated with the quasi-normality hypothesis.

\footnotetext{Though the DIA had yielded several important insights into the dynamics underlying the turbulence problem, controversies persist about the comparison of the predictions of the DIA with experiments at high Reynolds numbers (Mou and Weichman \cite{weichman}, Eyink \cite{eyink})}

The formal application of the DIA to a statistical problem is typically valid when the non-linear effects are weak.  Since these effects are not weak in the turbulence problem, there is a need to rationalize the DIA prior to application to this problem.  Kraichnan \cite{Kraichnan-1965} sought to do this via a model equation which can be solved exactly.  After applying the DIA to the model equation, it is modified by a process, where the non-linear terms are unrestricted, into another equation for which the DIA gives the exact statistical average.

Several mathematical issues associated with the application of the DIA to Kraichnan's \cite{Kraichnan-1961} stochastic model equation were explored by Shivamoggi, et al. \cite{Sen-Paper}.  This model equation is based on a Markovian process.  In this paper, we consider instead a stochastic model equation based on a \textit{non-Markovian} process (Zwanzig \cite{Zwanzig}), and explore the concomitant mathematical issues.

The DIA procedure is then applied to the problem of wave propagation in a random medium - this exploration is of relevance to the propagation of radio waves through the ionosphere and laser beams through the atmosphere (Tatarskii \cite{Tatarskii}, Chernov \cite{Chernov})\footnotemark.  This problem is described by a stochastic differential equation with the characteristics of the medium represented by stochastic coefficients (Bourret \cite{Bourret}, Keller \cite{Keller}, van Kampen \cite{vanKampen}, Shivamoggi et al. \cite{Sen2}),
\begin{equation}
	\frac{d^2E}{d\xi^2} + k^2[1 + \mu(\xi)]E = 0
\end{equation}
where $E$ is the electric field of a monochromatic wave, $k$ is the wavenumber of the wave, $\mu(\xi)$ represents the fluctuations in the refractive index of the medium (which is assumed to be a centered random function of position), and $\xi$ is the distance measured along the propagation direction.

\footnotetext{Although the fluctuations in the refractive index $\mu(\xi)$ of the medium in a turbulent atmosphere are very small, a wave propagates through a large number of  refractive index inhomogeneities in a typical situation of practical interest, so the cumulative effect can be very significant.}

\subsection*{2. The Non-Markovian Stochastic Model Equation}
Consider the stochastic equation (\cite{Zwanzig}):
\begin{subequations} \label{eqn:main}
	\begin{align}
		[\frac{d}{dt} + ib(t)]\hat G(t) + \int_{0}^{t}\Gamma(t-t')\hat G(t')dt' = \delta(t),
	\end{align}
	where $\Gamma(t)$ is a \textit{history-dependent} damping coefficient,
	\begin{align}
		&\Gamma(t) = \nu e^{-\mu t}, \nu \text{ and } \mu > 0.
	\end{align}
\end{subequations}

In the limit $\mu \rightarrow \infty$, the damping process becomes Markovian, first explored by Kraichnan \cite{Kraichnan-1961}.  On the other hand, in the limit $\mu \rightarrow 0$, this process has infinite memory and becomes ultra non-Markovian; eqn\eqref{eqn:main} then leads to:
\begin{equation}\label{eqn:main-2}
	[\frac{d}{dt} + ib(t)]\hat G(t) + \nu \int_{0}^{t}\hat G(t')dt' = \delta(t).
\end{equation}

Also, we assume that $b(t)$ is a real, centered, stationary Gaussian random function of $t$ described by the Uhlenbeck-Ornstein model \cite{Uhlenbeck-O} for the autocorrelation function of $b(t)$,
\begin{equation} \label{eqn:autocorrelation}
	\langle b(t)b(t') \rangle = \sigma^2e^{-\lambda(t-t')}.
\end{equation}

\subsubsection*{(i) The Perturbative Solution}

We first apply Keller's perturbative procedure \cite{Keller} to solve eqn\eqref{eqn:main-2}.  For this purpose, we put
\begin{subequations}
	\begin{align}
		L_0 \equiv D + \nu D^{-1},\, L_1 \equiv ib(t).
	\end{align}
	Eqn\eqref{eqn:main-2} then leads to
	\begin{align}
		L_0G(t) - \langle L_1 L_0^{-1} L_1 \rangle G(t) = \delta(t), \>\> G(t) \equiv \langle \hat G(t) \rangle.
	\end{align}
\end{subequations}

Now, noting that
\begin{equation}
[D^2 + \nu]h = (D +\nu D^{-1})Dh = \delta(t)
\end{equation}
the Green's function for $L_0$ is given by
\begin{equation}
	Dh = cos \sqrt{\nu}(t-t').
\end{equation}
Using (7), eqn (5b) becomes
\hspace{-3pt}
\begin{equation} \label{eqn:G-integral}
	\frac{dG}{dt} + \nu\! \int_0^t \!\!G(t')dt' - \!\int_0^t \!\langle ib(t)ib(t') \rangle cos\sqrt{\nu}(t-t')G(t')dt' = \delta(t).
\end{equation}

Using the Uhlenbeck-Ornstein model (\ref{eqn:autocorrelation}), eqn(\ref{eqn:G-integral}) becomes
\hspace{-3pt}
\begin{equation} \label{eqn:long-integral}
	\frac{dG}{dt} + \nu\! \int_0^t \!\!G(t')dt' + \sigma^2 \!\int_0^t e^{-\lambda(t-t')} cos\sqrt{\nu}(t-t')G(t')dt' = \delta(t).
\end{equation}

Upon taking the Laplace transform with respect to $t$, eqn(\ref{eqn:long-integral}) leads to
\begin{equation} \label{eqn:perturb-Laplace}
	\mathscr{G}(p) \left[p + \frac{\nu}{p} + \sigma^2\frac{p+\lambda}{(p+\lambda)^2 + \nu}\right] = 1
\end{equation}
where,
\begin{equation*}
	\mathscr{G}(p) \equiv \int_{0}^{\infty} e^{-pt}G(t)dt.
\end{equation*}

Assuming that $\lambda$, $\sigma^2$, and $\nu$ are small quantities, we discard terms that are quadratic and higher powers in these quantities and obtain

\begin{equation}
	\mathscr{G}(p-\lambda) \approx \frac{p^2 - \lambda p + \nu}{p\,(p^2 - 2\lambda p + 2\nu + \sigma^2)}.
\end{equation}

Upon doing the partial fraction decomposition,

\begin{equation}
	\mathscr{G}(p-\lambda) \approx \frac{\dfrac{\nu}{2\nu + \sigma^2}}{p} + \frac{\dfrac{\nu + \sigma^2}{2\nu + \sigma^2}(p-\lambda)}{(p-\lambda)^2 + 2\nu + \sigma^2}  
\end{equation}
and inverting the Laplace transform, we obtain
\begin{equation}\label{eqn:perturb}
	G(t) \approx \frac{\nu}{2\nu + \sigma^2}e^{-\lambda t} + \frac{\nu + \sigma^2}{2\nu + \sigma^2}cos\sqrt{2\nu + \sigma^2}t
\end{equation}
which, in the limit $t \rightarrow \infty$, is seen to oscillate indefinitely.

\subsubsection*{(ii) The DIA Solution}
Application of the DIA procedure entails replacing the perturbative expression for the Green's function by the exact expression $G(t-t')$ in eqn(\ref{eqn:G-integral}), which yields

\begin{equation} \label{eqn:14}
	\frac{dG}{dt} + \nu\! \int_0^t \!\!G(t')dt' - \!\int_0^t \!\langle ib(t)ib(t') G(t-t') \rangle G(t')dt' = \delta(t).
\end{equation}
Using the \textit{weak-statistical dependence} hypothesis (Kraichnan \cite{K-1958}, \cite{K-1959}), we have
\begin{equation}\label{eqn:quasi-normality}
	\langle b(t)b(t')\hat G(t-t') \rangle = \langle b(t)b(t') \rangle \langle \hat G(t-t')\rangle.
\end{equation}

Using (\ref{eqn:quasi-normality}) and the Uhlenbeck-Ornstein model (\ref{eqn:autocorrelation}), and taking the Laplace transform with respect to $t$, eqn(\ref{eqn:14}) leads to a continued fraction solution for $\mathscr{G}$,
\begin{equation}\label{eqn:cont-frac}
	\mathscr{G}(p) = \frac{1}{p + \dfrac{\nu}{p} + \sigma^2\mathscr{G}(p+\lambda)}.
\end{equation}

Eqn\eqref{eqn:cont-frac} implies that $G(t)$ is of an exponential order; the appearance of $\mathscr{G}(p+\lambda)$ on the right implies, on analytically continuing $\mathscr{G}(p)$ into the left half of the complex $p$-plane, that $G(t)$ exhibits a time-dependence like $e^{-\lambda t}$.\footnotemark  \hspace{8pt}This is confirmed in the following (see \eqref{eqn:23}).

\footnotetext{We are thankful to Professor Greg Eyink for this observation.}

\begin{equation}
	\mathscr{G}(p) = \cfrac{p}{[p^2 + \nu] + \cfrac{\sigma^2p(p+\lambda)}{[(p+\lambda)^2 + \nu] + \cfrac{\sigma^2(p+\lambda)(p+2\lambda)}{(p+2\lambda)^2 + \nu + ...}}}
\end{equation}
We now analyze the various approximants of this continued fraction.
\begin{itemize}
\item The first approximant is
\begin{equation}
	\mathscr{G}^{(1)}(p) = \frac{p}{p^2 + \nu} 
\end{equation}
which totally ignores the stochastic element in eqn\eqref{eqn:main}.
\item The second approximant is
\begin{equation}
	\mathscr{G}^{(2)}(p) = \cfrac{p}{p^2 + \nu + \cfrac{\sigma^2p(p+\lambda)}{(p+\lambda)^2 + \nu}}
\end{equation}
which represents the perturbative solution \eqref{eqn:perturb-Laplace}.
	\item The third approximant is
	\begin{equation} \label{eqn:third-approxim}
	\mathscr{G}^{(3)}(p) = \cfrac{p}{[p^2 + \nu] + \cfrac{\sigma^2p(p+\lambda)}{[(p+\lambda)^2 + \nu] + \cfrac{\sigma^2(p+\lambda)(p+2\lambda)}{(p+2\lambda)^2 + \nu}}}
\end{equation}
which captures the non-perturbative aspects omitted by the perturbative solution \eqref{eqn:perturb}.
\end{itemize}

Let us assume again that $\lambda$, $\nu$, and $\sigma^2$ are small, so we discard terms that are quadratic and higher in $\lambda$, $\nu$, and $\sigma^2$.  Then eqn(\ref{eqn:third-approxim}) may be rewritten as

\begin{equation}\label{eqn:third-approx}
	\mathscr{G}^{(3)}(p-\lambda) = \frac{p^2 + \lambda p + 2\nu + \sigma^2}{p^3 + (3\nu+2\sigma^2)p}.
\end{equation}

Upon doing the partial fraction decomposition,

\begin{equation}
	\mathscr{G}^{(3)}(p-\lambda) \approx \frac{\dfrac{2\nu + \sigma^2}{3\nu + 2\sigma^2}}{p} + \left( \frac{\nu + \sigma^2}{3\nu + 2\sigma^2} \right)\frac{p}{p^2 + (3\nu + 2\sigma^2)} + \frac{\lambda}{p^2 + (3\nu + 2\sigma^2)}
\end{equation}
and inverting the Laplace transform, we obtain
\begin{equation} \label{eqn:23}
	G^{(3)}(t) \approx \frac{2\nu + \sigma^2}{3\nu + 2\sigma^2}e^{-\lambda t} + \frac{\nu + \sigma^2}{3\nu + 2\sigma^2}e^{-\lambda t}cos\sqrt{3\nu + 2\sigma^2}t \\
	+ \frac{\lambda}{\sqrt{3\nu + 2\sigma^2}}e^{-\lambda t}sin\sqrt{3\nu + 2\sigma^2}t
\end{equation}
which, in the limit $t \rightarrow \infty$, decays, as predicted previously, unlike the perturbative solution (\ref{eqn:perturb}).

It may be mentioned that alternative methods of solution of eqn(3) exist, which entail transforming eqn(3) into a second order differential equation, simplifying this equation using a Liouville transformation, and constructing several approximate solutions, with formally improving accuracy (see Appendix).

\subsection*{3. Wave Propagation Problem in a Random Medium}
\subsubsection*{(i) The Perturbative Solution}

Here we apply Keller's perturbation procedure \cite{Keller} to the second order equation governing the wave propagation in a random medium,

\begin{equation}\label{eqn:wave-prop-model}
	\left(\frac{d^2}{d\xi^2} + k^2\right)E +  \, k^2 \, \mu(\xi)E = \delta(\xi).
\end{equation}
where $\mu(\xi)$ represents the refractive index fluctuations in the medium.  Introducing the operators,
\begin{equation}
	L_0 \equiv \frac{d^2}{d\xi^2} + k^2, \, L_1 \equiv k^2 \, \mu(\xi)
\end{equation}
and applying Keller's perturbative procedure \cite{Keller}, eqn(\ref{eqn:wave-prop-model}) leads to
\begin{equation}\label{eqn:wave2}
	\left(\frac{d^2}{d\xi^2} + k^2 \right)\langle E \rangle \approx \int_0^\xi \langle k^2\mu(\xi)k^2\mu(\eta)\rangle \frac{1}{k}\sin{k(\xi - \eta)} \langle E(\eta) \rangle d\eta + \delta(\xi)
\end{equation}
where the Green's function for $L_0$ is given by
\begin{equation}
	G(\xi, \eta) = \frac{1}{k}\sin{k(\xi - \eta)}.
\end{equation}

Assuming the Uhlenbeck-Ornstein stochastic model for the refractive index fluctuations $\mu(\xi)$,
\begin{equation}\label{eqn:wave-autocorrelation}
	\langle \mu(\xi) \mu(\eta) \rangle = \sigma^2 e^{-\lambda(\xi - \eta)}
\end{equation}
eqn(\ref{eqn:wave2}) becomes
\begin{equation}\label{eqn:wave-prop-model2}
	\left(\frac{d^2}{d\xi^2} + k^2 \right)\langle E(\xi) \rangle = k^3 \sigma^2 \int_0^\xi e^{-\lambda(\xi - \eta)} \sin{k(\xi - \eta)} \langle E(\eta) \rangle d\eta + \delta(\xi).
\end{equation}

Laplace transforming with respect to $\xi$, we obtain from eqn(\ref{eqn:wave-prop-model2}),

\begin{equation}
	\mathscr{E}(p) = \frac{1}{p^2 + k^2 - \dfrac{k^4 \sigma^2}{(p + \lambda)^2 + k^2}}.
\end{equation}

Considering the non-Markovian limit ($\lambda$ small) and the geometrical optics limit (k large), and using the binomial theorem, we obtain
\begin{spreadlines}{0.8em}
\begin{align*}
	\mathscr{E}(p) &= \frac{1}{(p^2 + k^2) - \dfrac{k^4 \sigma^2}{k^2\left[\dfrac{(p+\lambda)^2}{k^2} + 1 \right]}} \\
	&\approx \frac{1}{(p^2 + k^2) - k^2\sigma^2 \left[1 - \dfrac{(p+\lambda)^2}{k^2}\right]}
\end{align*}
or
\begin{align}
	\mathscr{E}(p) &\approx \frac{1}{(1 + \sigma^2)p^2 + 2\sigma^2\lambda p + (1 - \sigma^2)k^2} \nonumber \\
	&\approx \frac{\dfrac{1}{1 + \sigma^2}}{\left(p + \dfrac{\sigma^2\lambda}{1 + \sigma^2} \right)^2 + \dfrac{k^2(1 - \sigma^4)}{(1 + \sigma^2)^2}}. \label{eqn:E-simple}
\end{align}
\end{spreadlines}

Upon inverting the Laplace transform, (\ref{eqn:E-simple}) leads to
\begin{equation} \label{eqn:last-eqn}
	E(\xi) \approx e ^{- \dfrac{\sigma^2\lambda}{1 + \sigma^2} \xi} \sin{\sqrt{\frac{1 - \sigma^2}{1 + \sigma^2}}k\xi}.
\end{equation}
Eqn(\ref{eqn:last-eqn}) shows the attenuation (\textit{fluctuation-dissipation theorem}, Huang \cite{Huang}) of the coherent wave due to refractive index fluctuations in the medium.  The latter also introduce a wavenumber shift.

\subsubsection*{(ii) The DIA Solution}
Application of the DIA procedure involves replacing the perturbative expression for the Green's function by the exact expression $E(\xi-\eta)$ in eqn(\ref{eqn:wave2}), which leads to
\begin{equation} \label{eqn:wave-prop-dia1}
	\left[ \frac{d^2}{d\xi^2} + k^2 \right]\langle E(\xi) \rangle \approx \int_0^{\xi} \langle k^2\mu(\xi)k^2\mu(\eta) E(\xi-\eta) \rangle \langle E(\eta) \rangle d\eta + \delta(\xi).
\end{equation}

Using (\ref{eqn:quasi-normality}) and the Uhlenbeck-Ornstein model (\ref{eqn:autocorrelation}), eqn(\ref{eqn:wave-prop-dia1}) leads to
\begin{equation}\label{eqn:wave-prop-dia}
	\left[ \frac{d^2}{d\xi^2} + k^2 \right]\langle E(\xi) \rangle - \sigma^2 k^4 \int_0^{\xi} e^{-\lambda (\xi - \eta)} \langle E(\xi-\eta) \rangle \langle E(\eta) \rangle d\eta = \delta(\xi).
\end{equation}

Upon Laplace transforming with respect to $\xi$, eqn(\ref{eqn:wave-prop-dia}) leads to the functional equation,

\begin{equation}\label{eqn:functional}
	(p^2 + k^2)\mathscr{E}(p) - \sigma^2k^4\mathscr{E}(p+\lambda)\mathscr{E}(p) = 1.
\end{equation}
Eqn(\ref{eqn:functional}) has the continued fraction solution,

\begin{equation}\label{eqn:wave-cont-frac}
	\mathscr{E}(p) = \frac{1}{(p^2 + k^2) - \dfrac{\sigma^2 k^4}{[(p+\lambda)^2 + k^2] - \dfrac{\sigma^2 k^4}{[(p + 2\lambda)^2 + k^2]+...}}}
\end{equation}

Successive truncations of (\ref{eqn:wave-cont-frac}) lead to the following approximants,

\begin{equation}\label{eqn:approximants}
	\mathscr{E}(p) = \frac{1}{p^2 + k^2},\, \frac{1}{p^2+k^2 - \dfrac{\sigma^2 k^4}{(p+\lambda)^2 + k^2}}, \, \frac{1}{(p^2 + k^2) - \dfrac{\sigma^2 k^4}{[(p+\lambda)^2 + k^2] - \dfrac{\sigma^2 k^4}{[(p+ 2\lambda)^2 + k^2]}}}
\end{equation}

Note that while the second approximant in (\ref{eqn:approximants}) corresponds to the perturbative solution, the third approximant coresponds to the non-perturbative solution.  Note the latter can be approximated in the limit, small $\lambda$ and large $k$, as before, by

\begin{align} \label{eqn:E}
	\mathscr{E}(p) &\approx \frac{1}{(p^2 + k^2) - \dfrac{\dfrac{\sigma^2 k^4}{(1+\sigma^2)}}{\left[ (p+\lambda+\dfrac{\sigma^2 \lambda}{1 + \sigma^2})^2 + k^2 (\dfrac{1 - \sigma^2}{1 + \sigma^2}) \right]}}.
\end{align}

Putting,
\begin{equation} \label{eqn:defining-bits}
	\alpha \equiv \frac{1 - \sigma^2}{1 + \sigma^2}, \,
	\hat \lambda \equiv \frac{1 + 2\sigma^2}{1 + \sigma^2} \lambda, \,
	\beta \equiv \frac{\sigma^2}{1 - \sigma^2},
\end{equation}
(\ref{eqn:E}) may be rewritten as
\begin{align*}
	\mathscr{E}(p) &\approx \frac{1}{(p^2 + k^2) - \dfrac{\sigma^2 k^2}{1 - \sigma^2}\left[1 - \dfrac{(p+ \hat \lambda)^2}{k^2 \alpha} \right]}
\end{align*}
or
\begin{equation} \label{eqn:laplace-final-wave}
	\mathscr{E}(p) \approx \frac{\dfrac{1}{(1+\dfrac{\beta}{\alpha})}}{\left(p + \dfrac{\dfrac{\beta}{\alpha}}{1 + \dfrac{\beta}{\alpha}}\hat\lambda \right)^2 + \dfrac{k^2(1 - \beta)}{1 + \dfrac{\beta}{\alpha}}}.
\end{equation}

Inverting the Laplace transform, \eqref{eqn:laplace-final-wave} leads to

\begin{equation} \label{eqn:wave-dia-soln}
	E(\xi) \approx e^{-\dfrac{\sigma^2(1+2\sigma^2)}{1 - \sigma^2 + 2\sigma^4}\lambda\xi} \sin{\sqrt{\frac{1-3\sigma^2+2\sigma^4}{1-\sigma^2+2\sigma^4}}k\xi}.
\end{equation}

Noting that the DIA wave attenuation coefficient is given by
\begin{equation}
	\dfrac{\sigma^2(1+2\sigma^2)}{1 - \sigma^2 + 2\sigma^4} = (\frac{\sigma^2}{1 + \sigma^2})(\frac{1 + 3\sigma^2 + 2\sigma^4}{1 - \sigma^2+2\sigma^4}) > \frac{\sigma^2}{1 + \sigma^2}
\end{equation}
one notices on comparison with the perturbative solution (\ref{eqn:last-eqn}), that the latter underestimates the wave attenuation in a random medium. Further noting that,
\begin{equation}
	\frac{1-3\sigma^2+2\sigma^4}{1-\sigma^2+2\sigma^4} = \frac{1 - \sigma^2}{1 + \sigma^2} - \frac{4\sigma^4(1-\sigma^2)}{1 - 2\sigma^2(1-\sigma^2)}
\end{equation}
one notices, on comparison with the perturbative solution (\ref{eqn:last-eqn}), that the latter provides a good estimate of the wavenumber shift produced by the refractive index fluctuations in a random medium.

\subsection*{4. Discussion}
Thanks to the requirement of weak nonlinear effects for formal application of the DIA to a statistical problem, a rationalization of the DIA is in order whenever such a requirement is violated, as in the turbulence problem.  In this paper, we have explored mathematical aspects associated with the application of the DIA to non-Markovianized stochastic models in the turbulence problem.  This process is shown to lead to a functional equation, and construction of solutions of this equation is formulated within the framework of a continued fraction representation.  The perturbative solution is shown to correspond to the second approximant of the latter representation, while the non-perturbative renormalization aspects are incorporated in the third (and higher) approximants.

The DIA procedure is next applied to the problem of wave propagation in a random medium, which is described by a stochastic differential equation, with the characteristics of the medium represented by stochastic coefficients. The perturbative solution is shown to provide a good estimate of the wavenumber shift produced by the refractive index fluctuations, but to underestimate the wave attenuation in a random medium.


\begin{appendices}
\setcounter{equation}{0}
\renewcommand{\theequation}{A.\arabic{equation}}
\subsection*{Appendix: Alternative Methods of Solution}

\subsection*{A.1 Second-order Differential Equation for the Non-Markovian \\ Stochastic Model Equation}
In this section we transform the model eqn\eqref{eqn:main-2} into a second order differential equation, simplify this equation using a Liouville transformation, and construct three approximate solutions, with formally successively improving accuracy.

Putting,
\begin{equation}\label{eqn:Ghat-defn}
	\hat G(t) \equiv \frac{dg}{dt}
\end{equation}
eqn\eqref{eqn:main-2} becomes
\begin{equation}\label{eqn:second-order}
	\frac{d^2g}{dt^2} + ib(t)\frac{dg}{dt} + \nu g = \delta(t).
\end{equation}

Making the Liouville transformation,
\begin{equation}\label{eqn:defn-g}
	g(t) = e^{-\frac{i}{2}\int_0^tb(t')dt'} f(t)
\end{equation}
eqn(\ref{eqn:second-order}) becomes
\begin{equation}\label{eqn:second-order-f}
	f''(t) + \left(\frac{1}{4}b^2(t) - \frac{i}{2}b'(t) + \nu \right)f(t) = \delta(t).
\end{equation}
Treating $b(t)$ as small, eqn(\ref{eqn:second-order-f}) becomes
\begin{equation}\label{eqn:second-order model}
	f''(t) + [\nu - \frac{i}{2}b'(t)]f = \delta(t).
\end{equation}

Rewriting eqn(\ref{eqn:second-order model}) as
\begin{equation}\label{eqn:second-order-2}
	\left(\frac{d^2}{dt^2} + \nu \right)f(t) - \frac{i}{2}b'(t)f(t) = \delta(t)
\end{equation}
and introducing the operators,
\begin{equation}
L_0 \equiv \frac{d^2}{dt^2} + \nu, \, L_1 \equiv - \frac{i}{2}b'(t)
\end{equation}
and applying Keller's perturbative procedure \cite{Keller}, eqn(\ref{eqn:second-order-2}) leads to
\begin{equation}\label{eqn:second-order model-1}
	\left(\frac{d^2}{dt^2} + \nu \right)\langle f(t) \rangle + \frac{1}{4} \int_0^t \langle b'(t)b'(t') \rangle \frac{1}{\sqrt{\nu}}\sin{\sqrt{\nu}(t - t')} \langle f(t') \rangle dt' = \delta(t)
\end{equation}
where the Green's function for $L_0$ is given by
 \begin{equation}
 	G(t, t') = \frac{1}{\sqrt{\nu}}\sin{\sqrt{\nu}(t - t')}.
 \end{equation}

Introducing the \textit{centroid} coordinates,
\begin{equation}\label{eqn:centroid}
		T \equiv \frac{1}{2}(t + t'), \, \tau \equiv t - t'
\end{equation}
and using the Uhlenbeck-Ornstein model (\ref{eqn:autocorrelation}), we have
\begin{equation}
\langle b'(t)b'(t') \rangle = \frac{\partial^2}{\partial t \partial t'} \langle b(t)b(t') \rangle = \left(\frac{1}{2}\frac{\partial}{\partial T} + \frac{\partial}{\partial \tau} \right) \left( \frac{1}{2} \frac{\partial}{\partial T} - \frac{\partial}{\partial \tau} \right) \left[\sigma^2e^{-\lambda\tau} \right]
\end{equation}
from which,
\begin{equation}\label{eqn:autocorrelation-b-prime}
	\langle b'(t)b'(t') \rangle = -\sigma^2\lambda^2e^{-\lambda(t-t')},
\end{equation}

Using (\ref{eqn:autocorrelation-b-prime}), eqn(\ref{eqn:second-order model-1}) becomes
\begin{equation}
	\left(\frac{d^2}{dt^2} + \nu \right)\langle f(t) \rangle - \frac{\sigma^2\lambda^2}{4\sqrt{\nu}} \int_0^t e^{-\lambda(t-t')} \sin{\sqrt{\nu}(t - t')} \langle f(t') \rangle dt' = \delta(t).
\end{equation}
Upon doing the Laplace transform with respect to t, we obtain
\begin{equation}\label{eqn:model2-Laplace}
	F(p) = \frac{1}{p^2 + \nu - \dfrac{\dfrac{\sigma^2\lambda^2}{4}}{(p + \lambda)^2 + \nu}}.
\end{equation}

We now attempt to invert the Laplace transform given by (\ref{eqn:model2-Laplace}) in several ways with formally successively improving accuracy.

\subsubsection*{(i)  The first approximation}
As a first approximation, if we ignore $\lambda$ in the bracket in the denominator of (\ref{eqn:model2-Laplace}), we then have
\begin{equation}
		F(p) = \frac{1}{p^2 + k^2 - \dfrac{\alpha^2}{p^2 + k^2}},
\end{equation}
which may be reorganized as follows,
\begin{align*}
	F(p) &= \frac{p^2 + k^2}{(p^2 + k^2)^2 - \alpha^2} 
\end{align*}
or
\begin{equation} \label{eqn:laplace-simplified}
	F(p) = \frac{1}{2} \left( \frac{1}{p^2 + k^2 + \alpha} + \frac{1}{p^2 + k^2 - \alpha} \right)
\end{equation}
where, $$ \alpha^2 \equiv \frac{\sigma^2\lambda^2}{4} \text{ and } k^2 \equiv \nu.$$

(\ref{eqn:laplace-simplified}) may be easily inverted to give
\begin{equation}\label{eqn:f-1}
	\langle f(t) \rangle = \frac{1}{2\sqrt{k^2 + \alpha}} \sin{\sqrt{k^2 + \alpha}t} + \frac{1}{2\sqrt{k^2 - \alpha}} \sin{\sqrt{k^2 - \alpha}t}.
\end{equation}

Noting,
\begin{align*}
 \sqrt{k^2 + \alpha} = k\sqrt{1 + \dfrac{\alpha}{k^2}} \approx k[1+ \dfrac{\alpha}{2k^2}] \\
 \sqrt{k^2 - \alpha} = k\sqrt{1 - \dfrac{\alpha}{k^2}} \approx k[1 - \dfrac{\alpha}{2k^2}],
\end{align*}
(\ref{eqn:f-1}) becomes
\begin{equation}
	\begin{aligned}
		\langle f(t) \rangle &\approx \frac{1}{2} \left[ \frac{1}{k}(1 - \frac{\alpha}{2k^2} )\sin{k(1 + \frac{\alpha}{2k^2})t} + \frac{1}{k}(1 + \frac{\alpha}{2k^2})\sin{k(1 - \frac{\alpha}{2k^2})t} \right], \\
		\text{or} \\
		\langle f(t) \rangle &\approx \frac{1}{2k} \bigg[(1 - \frac{\alpha}{2k^2})[\sin{kt} \cos{\frac{\alpha}{2k}t} + \cos{kt} \sin{\frac{\alpha}{2k}t}] + (1 + \frac{\alpha}{2k^2})[\sin{kt}\cos{\frac{\alpha}{2k}t} \\ &- \cos{kt} \sin{\frac{\alpha}{2k}t}] \bigg].
	\end{aligned}
\end{equation}
Approximating further,
\begin{align*}
	\cos{\frac{\alpha t}{2k}} &\approx 1 - \frac{\alpha^2t^2}{8k^2} \\
	\sin{\frac{\alpha t}{2k}} &\approx \frac{\alpha}{2k}t
\end{align*}
and discarding terms $O(\alpha^3)$, we have
\begin{equation}
	\begin{aligned}
	\langle f(t) \rangle \approx &\frac{1}{2k} \bigg[(1 - \frac{\alpha}{2k^2}) [(1 - \frac{\alpha^2t^2}{8k^2})\sin{kt} + \frac{\alpha t}{2k}\cos{kt}] + (1 + \frac{\alpha}{2k^2}) [(1 - \frac{\alpha^2t^2}{8k^2}) \sin{kt} \\ &- \frac{\alpha t}{2k} \cos{kt}] \bigg],
\end{aligned}
\end{equation}
which may be reorganized as follows,

\begin{align*}
	\langle f(t) \rangle &= \frac{1}{k} \left[\sin{kt} - \frac{\alpha^2t^2}{8k^2}\sin{kt} - \frac{\alpha^2t}{4k^3}\cos{kt} \right] \\
	&\approx \frac{1}{k} \left[\sin{kt}\cos{\frac{\alpha^2t}{4k^3}} - \sin{\frac{\alpha^2t}{4k^3}}\cos{kt} - \frac{\alpha^2t^2}{8k^2}\sin{kt} \right] \\
	&\approx \frac{1}{k} \left[\sin{k(1 - \frac{\alpha^2}{4k^4})t} - \frac{\alpha^2t^2}{8k^2}\sin{kt} \right] \\
	&\approx \frac{1}{k} \left[\sin{k(1 - \frac{\alpha^2}{4k^4})t} - \frac{\alpha^2t^2}{8k^2}\sin{k(1 - \frac{\alpha^2}{4k^4})t} \right] \\
	&\approx \frac{1}{k} e^{-\dfrac{\alpha^2t^2}{8k^2}} \sin{k(1 - \frac{\alpha^2}{4k^4})t} \\
	&= \frac{1}{k}e^{-\dfrac{\sigma^2\lambda^2}{32k^2}t^2} \sin{(1-\frac{\sigma^2\lambda^2}{16k^4})t} \numberthis \label{eqn:f}
\end{align*}

Using (\ref{eqn:f}), \eqref{eqn:defn-g} gives
\begin{equation}\label{eqn:g-formula}
	g(t) = \frac{1}{k} e^{-\frac{i}{2}\int_0^tb(t')dt'}  e^{-\dfrac{\sigma^2\lambda^2}{32k^2}t^2} \sin{k(1 - \frac{\sigma^2\lambda^2}{16k^4})t}
\end{equation}

Using (\ref{eqn:g-formula}), (\ref{eqn:Ghat-defn}) gives
\begin{equation}\label{eqn:avg-Ghat}
		\hat G(t) = \frac{dg}{dt} \approx e^{-\frac{i}{2}\int_0^tb(t')dt'}e^{-\dfrac{\sigma^2\lambda^2}{32k^2}t^2} \cos{k(1 - \frac{\sigma^2\lambda^2}{16k^4})t}
\end{equation}


Using (\ref{eqn:autocorrelation}), we have\footnotemark
\begin{equation}\label{eqn:avg-G}
	\langle e^{-\frac{i}{2}\int_0^tb(t')dt'} \rangle = e^{-\sfrac{1}{4} \int_0^t \int_0^t \langle b(t')b(t'') \rangle dt' dt''} =  e^{-\dfrac{\sigma^2}{4\lambda^2}(\lambda t - 1 + e^{-\lambda t})} =
	\begin{cases}
		e^{-\dfrac{\sigma^2}{8}t^2} \text{, }\lambda \text{ small} \\
		e^{-\dfrac{\sigma^2}{4\lambda}t} \text{, }\lambda \text{ large}
	\end{cases}.
\end{equation}

Using (\ref{eqn:avg-G}), in the small-$\lambda$ limit, (\ref{eqn:avg-Ghat}) leads to
\begin{equation}\label{eqn:f1-final}
		G(t) \equiv \langle \hat G(t) \rangle \approx e^{-(1+\dfrac{\lambda^2}{4k^2})\dfrac{\sigma^2}{8}t^2} \cos{(1-\frac{\sigma^2\lambda^2}{16k^4})kt}.
\end{equation}

\footnotetext{In evaluating the double integral, we have used the result, $\int_0^T\int_0^T \rho(t - t')dt dt' = 2 \int_0^T (1 - \sfrac{\tau}{T})\rho(\tau) d\tau.$}

\subsubsection*{(ii)  The second approximation}
Restoring $\lambda$ in the bracket in the denominator of (\ref{eqn:model2-Laplace}), and treating it as a small quantity, we obtain
\begin{equation}
	F(p) = \frac{1}{p^2 + k^2 - \dfrac{\dfrac{\sigma^2\lambda^2}{4}}{(p + \lambda)^2 + k^2}}
\end{equation}
which may be simplified as follows,
\begin{spreadlines}{0.8em}
\begin{align}
	F(p) &= \frac{1}{p^2 + k^2 - \dfrac{\dfrac{\sigma^2\lambda^2}{4}}{p^2\left(1 + \dfrac{\lambda}{p}\right)^2 + k^2}} \nonumber
	\\
	&\approx \frac{p^2 + k^2 + 2\lambda p}{(p^2 + k^2)^2 + 2\lambda p(p^2 + k^2) - \dfrac{\sigma^2\lambda^2}{4}} \nonumber
	\\
	&\approx \frac{p^2 + k^2 + 2\lambda p}{(p^2 + k^2 + \lambda p)^2 - \dfrac{\sigma^2\lambda^2}{4}} \nonumber \\
	&= \frac{(p^2 + k^2 + \lambda p) + \lambda p}{(p^2 + k^2 + \lambda p + \dfrac{\sigma \lambda}{2})(p^2 + k^2 + \lambda p - \dfrac{\sigma \lambda}{2})}. \label{eqn:F}
\end{align}
\end{spreadlines}

(\ref{eqn:F}) may be decomposed into partial fractions as follows,

\begin{align}
	F(p)= &\frac{1}{2} \left[\frac{1}{p^2 + k^2 + \lambda p + \dfrac{\sigma \lambda}{2}} + \frac{1}{p^2 + k^2 + \lambda p - \dfrac{\sigma \lambda}{2}} \right] \nonumber
	\\ &- \frac{1}{\sigma} \left[\frac{p}{p^2 + k^2 + \lambda p + \dfrac{\sigma \lambda}{2}} - \frac{p}{p^2 + k^2 + \lambda p - \dfrac{\sigma \lambda}{2}} \right] \nonumber
\end{align}
or
\begin{align}
	F(p) \approx &\frac{1}{2} \left[\frac{1}{(p + \dfrac{\lambda}{2})^2 + (k^2 + \dfrac{\sigma \lambda}{2})} + \frac{1}{(p + \dfrac{\lambda}{2})^2 + (k^2 - \dfrac{\sigma \lambda}{2})} \right] \nonumber \\ &- \frac{1}{\sigma} \left[\frac{(p + \dfrac{\lambda}{2}) - \dfrac{\lambda}{2}}{(p + \dfrac{\lambda}{2})^2 + (k^2 + \dfrac{\sigma \lambda}{2})} - \frac{(p + \dfrac{\lambda}{2}) - \dfrac{\lambda}{2}}{(p + \dfrac{\lambda}{2})^2 + (k^2 - \dfrac{\sigma \lambda}{2})} \right].
\end{align}

Hence,
\begin{equation}\label{eqn:F-simple}
	F(p - \frac{\lambda}{2}) = \frac{1}{2} \left[\frac{1 + \dfrac{\lambda}{\sigma}}{p^2 + (k^2 + \dfrac{\sigma \lambda}{2})} + \frac{1 - \dfrac{\lambda}{\sigma}}{p^2 + (k^2 - \dfrac{\sigma \lambda}{2})} \right] - \frac{1}{\sigma} \left[\frac{p}{p^2 + (k^2 + \dfrac{\sigma \lambda}{2})} - \frac{p}{p^2 + (k^2 - \dfrac{\sigma \lambda}{2})} \right]
\end{equation}

Defining
\begin{equation}\label{eqn:defn:f}
	\langle f(t) \rangle \equiv e^{-\dfrac{\lambda t}{2}}l(t)
\end{equation}
and inverting the Laplace transform, (\ref{eqn:F-simple}) leads to
\begin{align}\label{eqn:l}
	l(t) = &\frac{1}{2} \left[\frac{1 + \dfrac{\lambda}{\sigma}}{\sqrt{k^2 + \dfrac{\sigma \lambda}{2}}} \sin{(\sqrt{k^2 + \dfrac{\sigma \lambda}{2}}\,t)} + \frac{1 - \dfrac{\lambda}{\sigma}}{\sqrt{k^2 - \dfrac{\sigma \lambda}{2}}}\sin{(\sqrt{k^2 - \dfrac{\sigma \lambda}{2}}\,t)} \right] \nonumber \\
	&-\frac{1}{\sigma} \left[\cos{(\sqrt{k^2 + \dfrac{\sigma \lambda}{2}}\,t)} - \cos{(\sqrt{k^2 - \dfrac{\sigma \lambda}{2}}\,t)} \right]
\end{align}

Noting,
\begin{align*}
	&\sqrt{k^2 + \dfrac{\sigma \lambda}{2}} = k\sqrt{1 + \dfrac{\sigma \lambda}{2k^2}} \approx k[1+ \dfrac{\sigma \lambda}{4k^2}] \\
	&\sqrt{k^2 - \dfrac{\sigma \lambda}{2}} = k\sqrt{1 - \dfrac{\sigma \lambda}{2k^2}} \approx k[1 - \dfrac{\sigma \lambda}{4k^2}] \numberthis\\
 	&\cos{\frac{\sigma \lambda t}{4k}} \approx 1 - \frac{\sigma^2 \lambda^2 t^2}{32k^2} \\
 	&\sin{\frac{\sigma \lambda t}{4k}} \approx \frac{\sigma \lambda}{4k}t
\end{align*}
eqn(\ref{eqn:l}) may be reorganized as follows,
\begin{align*}
	l(t) &\approx \frac{1}{2} \left[\frac{1}{k} (1 + \frac{\lambda}{\sigma}) (1 - \frac{\sigma\lambda}{4k^2}) \sin{kt(1 + \frac{\sigma \lambda}{4k^2})} + \frac{1}{k} (1 - \frac{\lambda}{\sigma})(1 + \frac{\sigma\lambda}{4k^2}) \sin{kt(1 - \frac{\sigma\lambda}{4k^2})} \right] \\
	& -\frac{1}{\sigma} \left[\cos{kt(1 + \frac{\sigma\lambda}{4k^2})} - \cos{kt(1 - \frac{\sigma\lambda}{4k^2})} \right]
	\\
	& \approx \frac{1}{2k} \bigg\{(1 + \frac{\lambda}{\sigma} - \frac{\sigma\lambda}{4k^2} - \frac{\lambda^2}{4k^2})\left[(1 - \frac{\sigma^2 \lambda^2}{32k^2}t^2)\sin{kt} + \frac{\sigma \lambda t}{4k}\cos{kt}\right] \\ & + (1 - \frac{\lambda}{\sigma} + \frac{\sigma\lambda}{4k^2} - \frac{\lambda^2}{4k^2}) \left[(1 - \frac{\sigma^2\lambda^2 }{32k^2}t^2)\sin{kt} - \frac{\sigma\lambda t}{4k} \cos{kt} \right] \bigg\} + \frac{1}{\sigma} \left[\frac{2\sigma\lambda}{4k}t\sin{kt} \right]
	\\
	&\approx \frac{1}{k}(1 - \frac{\lambda^2}{4k^2})(1 - \frac{\sigma^2\lambda^2}{32k^2}t^2) \sin{kt} + (1 - \frac{\sigma^2}{4k^2})\frac{\lambda^2}{4k^2}t\cos{kt} + \frac{2}{\sigma}\sin{kt}\sin{\frac{\sigma\lambda}{4k^2}kt}.\numberthis \label{eqn:l-simplified}
\end{align*}

Now recall that $\lambda$ is a small quantity.  Since we are working with an exact result only to $O(\lambda^2)$, we may introduce higher powers of $\lambda$ into our equation to facilitate simplification without compromising the accuracy to this order.  Thus, eqn(\ref{eqn:l-simplified}) may be rewritten as
\begin{multline}
	l(t) \approx \frac{1}{k}(1 - \frac{\lambda^2}{4k^2}) \left[ \sin{kt} \cos{[(1 - \frac{\sigma^2}{4k^2})\frac{\lambda^2}{4k^2} t]} + \sin{[(1 - \frac{\sigma^2}{4k^2})\frac{\lambda^2}{4k^2}t]} \cos{kt} \right] \\ - \frac{1}{k}(1 - \frac{\lambda^2}{4k^2}) \frac{\sigma^2\lambda^2}{32k^2}t^2 \sin{kt} + \frac{\lambda}{2k} t\sin{kt}
\end{multline}
which may be reorganized as follows,
\begin{equation}\label{eqn:l-simple}
	l(t) = \frac{1}{k}(1 - \frac{\lambda^2}{4k^2})\sin{[1 + \frac{\lambda^2}{4k^2}(1 - \frac{\sigma^2}{4k^2})]kt} - \frac{1}{k}(1 - \frac{\lambda^2}{4k^2}) \frac{\sigma^2 \lambda^2}{32k^2}t^2\sin{kt} + \frac{\lambda}{2k}t\sin{kt}.
\end{equation}

(\ref{eqn:l-simple}) may be rewritten further as

\begin{equation}\label{eqn:defn-l-2}
	l(t) \approx \frac{1}{k} (1 - \frac{\lambda^2}{4k^2}) e^{\dfrac{\lambda t}{2}-\dfrac{\sigma^2\lambda^2 t^2}{32k^2}}\sin{[1 + \frac{\lambda^2}{4k^2}(1 - \frac{\sigma^2}{4k^2})]kt} 
\end{equation}
for large $t$.

Using (\ref{eqn:defn-l-2}), (\ref{eqn:defn:f}) gives
\begin{subequations}
	\begin{equation}
		\langle f(t) \rangle \approx \frac{1}{k} (1 - \frac{\lambda^2}{4k^2}) e^{-\dfrac{\sigma^2\lambda^2}{32k^2}t^2}\sin{[1 + \frac{\lambda^2}{4k^2}(1 - \frac{\sigma^2}{4k^2})]kt},
	\end{equation}
or
	\begin{equation}\label{eqn:f2-final}
		\langle f(t) \rangle \approx \frac{1}{k} e^{-\dfrac{\sigma^2\lambda^2}{32k^2}t^2} \sin{(1 - \frac{\sigma^2\lambda^2}{16k^4})kt}.
	\end{equation}
\end{subequations}

Observe that (\ref{eqn:f2-final}) completely agrees with (\ref{eqn:f}), which appears to be reasonable since the first and second approximations agree to $O(\lambda^2)$.

\subsubsection*{(iii) The third approximation}
In the previous development, we dropped terms that were $O(\lambda^3)$, assuming they were very small.  The following calculations make no such assumption.


We may write (\ref{eqn:model2-Laplace}) as
\begin{equation}\label{eqn:Laplace-model-3}
	F(p - \lambda) = \frac{p^2 + k^2}{[(p - \lambda)^2 + k^2](p^2 + k^2) - \alpha^2}.
\end{equation}

Introducing

\begin{align}
	\langle f(t) \rangle &\equiv e^{-\lambda t}l(t) \nonumber \\
	\beta^2 &\equiv \alpha^2 - k^2\lambda^2 \equiv \mu^2\lambda^2 \label{eqn:defn-mu-squared} \\
	\mu^2 &\equiv \frac{\sigma^2}{4} - k^2, \nonumber
\end{align}
(\ref{eqn:Laplace-model-3}) leads to
\begin{equation}\label{eqn:L1}
	L(p) = \frac{p^2 + k^2}{(p^2 + k^2 - p\lambda)^2 - \beta^2}.
\end{equation}
(\ref{eqn:L1}) may be rewritten as,
\begin{equation}\label{eqn:L2}
	L(p) = \frac{(p^2 + k^2 -p\lambda) + p\lambda}{(p^2 + k^2 -p\lambda + \mu\lambda)(p^2 + k^2 - p\lambda - \mu \lambda)}.
\end{equation}

Upon doing the partial fraction decomposition, (\ref{eqn:L2}) becomes
\begin{subequations}
\begin{multline}
	L(p) = \frac{1}{2} \left[\frac{1}{p^2 + k^2 - p\lambda + \mu\lambda} + \frac{1}{p^2 + k^2 - p\lambda - \mu\lambda} \right] \\ - \frac{1}{2\mu\lambda} \left[\frac{p\lambda}{p^2 + k^2 -p\lambda + \mu\lambda} - \frac{p\lambda}{p^2 + k^2 - p\lambda -\mu\lambda} \right].
\end{multline}
which may be rewritten as,
\begin{multline} \label{eqn:L}
	L(p - \frac{\lambda}{2}) = \frac{1}{2}\left[ \frac{1}{(p-\dfrac{\lambda}{2})^2 + (k^2 + \mu\lambda - \dfrac{\lambda^2}{4})} + \frac{1}{(p-\dfrac{\lambda}{2})^2 + (k^2 - \mu\lambda - \dfrac{\lambda^2}{4})} \right] \\ - \frac{1}{2\mu\lambda} \left[ \frac{(p - \dfrac{\lambda}{2}) \lambda + \dfrac{\lambda^2}{2}}{(p-\dfrac{\lambda}{2})^2 + (k^2 + \mu\lambda - \dfrac{\lambda^2}{4})} - \frac{(p - \dfrac{\lambda}{2}) \lambda + \dfrac{\lambda^2}{2}}{(p-\dfrac{\lambda}{2})^2 + (k^2 - \mu\lambda - \dfrac{\lambda^2}{4})} \right].
\end{multline}

\end{subequations}

Upon introducing
\begin{equation}\label{eqn:defn-l}
l(t) \equiv e^{\dfrac{\lambda t}{2}}m(t)
\end{equation}
we obtain from (\ref{eqn:L})
\begin{multline}
	M(p) = \frac{1}{2} \left[\frac{1}{p^2 + (k^2 + \mu\lambda - \dfrac{\lambda^2}{4})} + \frac{1}{p^2 + (k^2 - \mu\lambda - \dfrac{\lambda^2}{4})} \right] \\ - \frac{1}{2\mu\lambda} \left[\frac{p\lambda + \dfrac{\lambda^2}{2}}{p^2 + (k^2 + \mu\lambda - \dfrac{\lambda^2}{4})} - \frac{p\lambda + \dfrac{\lambda^2}{2}}{p^2 + (k^2 - \mu\lambda - \dfrac{\lambda^2}{4})} \right].
\end{multline}

Inverting the Laplace transform, obtain
\begin{multline}\label{eqn:defn-m}
	m(t) = \frac{\dfrac{1}{2}(1 - \dfrac{\lambda}{2\mu})}{\sqrt{k^2 + \mu\lambda - \dfrac{\lambda^2}{4}}}\sin{\sqrt{k^2 + \mu\lambda - \dfrac{\lambda^2}{4}}\,t} + \frac{\dfrac{1}{2}(1 + \dfrac{\lambda}{2\mu})}{\sqrt{k^2 - \mu\lambda - \dfrac{\lambda^2}{4}}}\sin{\sqrt{k^2 - \mu\lambda - \dfrac{\lambda^2}{4}}\,t} \\ -
	\frac{1}{2\mu} \cos{\sqrt{k^2 + \mu\lambda - \dfrac{\lambda^2}{4}}\,t} + \frac{1}{2\mu} \cos{\sqrt{k^2 - \mu\lambda - \dfrac{\lambda^2}{4}}\,t}.
\end{multline}

Note,
\begin{equation} \label{eqn:simplifications}
	\begin{aligned}
		\sqrt{k^2 \pm \mu\lambda - \dfrac{\lambda^2}{4}} &\approx k \left[1 \pm \frac{\lambda\mu}{2k^2} - \lambda^2(\frac{k^2 + \mu^2}{8k^4}) \right] \\
		\frac{1}{\sqrt{k^2 \pm \mu\lambda - \dfrac{\lambda^2}{4}}} &\approx k \left[1 \mp \frac{\lambda\mu}{2k^2} + \lambda^2(\frac{k^2 + 3\mu^2}{8k^4}) \right].
	\end{aligned}
\end{equation}

Thus,
\begin{equation} \label{eqn:more-simplifications}
	\frac{\dfrac{1}{2}(1 \mp \dfrac{\lambda}{2\mu})}{\sqrt{k^2 \pm \mu\lambda - \dfrac{\lambda^2}{4}}} \approx \frac{1}{2k} \left[1 \mp \frac{\lambda}{2}(\frac{\mu}{k^2} + \frac{1}{\mu}) + \frac{3\lambda^2}{8}(\frac{k^2 + \mu^2}{k^4}) \right].
\end{equation}

Substituting (\ref{eqn:simplifications}) and (\ref{eqn:more-simplifications}) into (\ref{eqn:defn-m}), we obtain
\begin{multline}\label{eqn:m-long}
	m(t) = \frac{1}{2k}\left[1 - \frac{\lambda}{2} \left( \frac{\mu}{k^2} + \frac{1}{\mu} \right) + \frac{3\lambda^2}{8} \left( \frac{k^2+ \mu^2}{k^4} \right) \right] \cdot \bigg[ \sin{kt} \cos \{\frac{\lambda \mu}{2k^2} - \lambda^2 \bigg(\frac{k^2 + \mu^2}{8k^4}\bigg)\}kt
	\\ + \cos{kt} \sin{\{\frac{\lambda \mu }{2k^2} - \lambda^2 \bigg(\frac{k^2 + \mu^2}{8k^4}\bigg)\}kt} \bigg]
	\\ + \frac{1}{2k}\left[1 + \frac{\lambda}{2} \left( \frac{\mu}{k^2} + \frac{1}{\mu} \right) + \frac{3 \lambda^2}{8} \left( \frac{k^2+ \mu^2}{k^4} \right) \right] \cdot \bigg[ \sin{kt} \cos\{ \frac{\lambda \mu}{2k^2} + \lambda^2 \bigg(\frac{k^2 + \mu^2}{8k^4}\bigg)\}kt
 	\\ - \cos{kt} \sin{\{ \frac{\lambda \mu}{2k^2} + \lambda^2 \bigg(\frac{k^2 + \mu^2}{8k^4}\bigg)\}kt} \bigg]
	\\ - \frac{1}{2\mu} \bigg[ \cos{kt} \cos\{ \frac{\lambda \mu}{2k^2} - \lambda^2 \bigg(\frac{k^2 + \mu^2}{8k^4}\bigg)\}kt - \sin{kt} \sin\{ \frac{\lambda \mu}{2k^2} - \lambda^2 \bigg(\frac{k^2 + \mu^2}{8k^4}\bigg)\} kt \bigg]
	\\ + \frac{1}{2\mu} \bigg[ \cos{kt} \cos\{\frac{\lambda \mu}{2k^2} + \lambda^2 \bigg(\frac{k^2 + \mu^2}{8k^4}\bigg)\}kt + \sin{kt} \sin\{ \frac{\lambda \mu}{2k^2} + \lambda^2\bigg( \frac{k^2 + \mu^2}{8k^4}\bigg)\} kt \bigg].
\end{multline}

Utilizing the series approximations in powers of $\lambda$ for sine and cosine, and simplifying, (\ref{eqn:m-long}) becomes


\begin{equation}\label{eqn:m-simple1}
	m(t) \approx \frac{1}{k} (1 + \frac{\lambda}{2}t - \frac{\lambda^2 \mu^2 }{8k^2}t^2)\sin{kt} - \frac{1}{k} \frac{3\sigma^2}{32k^3}\lambda^2 t \cos{kt}
\end{equation}
which may be rewritten as,
\begin{equation}
	m(t) \approx \frac{1}{k} e^{ \dfrac{\lambda t}{2}  - \dfrac{\lambda^2\mu^2 t^2}{8k^2}} \sin (1 - \frac{3\sigma^2\lambda^2}{32k^4})kt.
\end{equation}

Returning to $l(t)$, as per (\ref{eqn:defn-l}), we have
\begin{equation}
	l(t) \approx \frac{1}{k} e^{\lambda t - \dfrac{\lambda^2\mu^2}{8k^2} t^2} \sin (1 - \frac{3\sigma^2\lambda^2}{32k^4})kt
\end{equation}
and thus, from (\ref{eqn:defn-mu-squared}),
\begin{equation} \label{eqn:f-avg}
	\langle f(t) \rangle \approx \frac{1}{k} e^{ - \dfrac{\lambda^2\mu^2}{8k^2} t^2} \sin (1 - \frac{3\sigma^2\lambda^2}{32k^4})kt.
\end{equation}

In the small-damping limit ($k$ small), eqn(\ref{eqn:f-avg}) becomes
\begin{equation}\label{eqn:f3-final}
	\langle f(t) \rangle \approx \frac{1}{k} e^{ - \dfrac{\lambda^2\sigma^2}{32k^2} t^2} \sin (1 - \frac{3\sigma^2\lambda^2}{32k^4})kt.
\end{equation}

(\ref{eqn:f3-final}) agrees with the previous results \eqref{eqn:f} and \eqref{eqn:f2-final} from the first and second approximations, respectively, except for the slight numerical discrepancy with the wavenumber shift.

\subsection*{A.2 First Order System for the Non-Markovian Stochastic Model Equation}
\label{sub:First Order System Approach}

We may write the second order equation (\ref{eqn:second-order-2}) as a system of first order equations by introducing,
\begin{equation}
	g \equiv \frac{df}{d\xi}, \, \xi \equiv \sqrt{\nu}t.
\end{equation}

Eqn(\ref{eqn:second-order-2}) is then equivalent to,
\begin{equation}\label{eqn:system}
	\begin{cases}
		\dfrac{df}{d\xi} &= g \\
		\dfrac{dg}{d\xi} &= -(1 - \dfrac{i}{2\nu}b'(\dfrac{\xi}{\sqrt{\nu}}))f.
	\end{cases}
\end{equation}

Applying Keller's \cite{Keller} perturbation procedure, eqn(\ref{eqn:system}) leads to the following matrix equation,

\begin{multline}\label{eqn:matrix-original}
	\frac{d}{d\xi} \begin{pmatrix} \langle f \rangle \\ \langle g \rangle \end{pmatrix} = \bigg[ \begin{pmatrix} 0 & 1 \\ -1 & 0 \end{pmatrix} \\ - \frac{1}{4\nu^2} \int_0^t \langle  b'(\dfrac{\xi}{\sqrt{\nu}}) b'(\dfrac{(\xi - \eta)}{\sqrt{\nu}}) \rangle \begin{pmatrix} 0 & 0 \\ -1 & 0 \end{pmatrix} e^{(\begin{smallmatrix} 0 & 1 \\ -1 & 0 \end{smallmatrix})\eta} \cdot \begin{pmatrix} 0 & 0 \\ -1 & 0 \end{pmatrix} e^{-(\begin{smallmatrix} 0 & -1 \\ 1 & 0 \end{smallmatrix})\eta} d\eta \bigg] \begin{pmatrix} \langle f \rangle \\ \langle g \rangle \end{pmatrix}
\end{multline}

Noting
\begin{equation}
	\begin{aligned}
		e^{(\begin{smallmatrix} 0 & 1 \\ -1 & 0 \end{smallmatrix})\eta} &= \begin{pmatrix} \cos{\eta} & \sin{\eta} \\ -\sin{\eta} & \cos{\eta} \end{pmatrix}, \\
	\end{aligned}
\end{equation}
eqn(\ref{eqn:matrix-original}) simplifies to
\begin{equation}\label{eqn:matrix-2}
	\frac{d}{d\xi} \begin{pmatrix} \langle f \rangle \\ \langle g \rangle \end{pmatrix} = \bigg[ \begin{pmatrix} 0 & 1 \\ -1 & 0 \end{pmatrix} - \frac{1}{4\nu^2} \int_0^t \langle  b'(\dfrac{\xi}{\sqrt{\nu}}) b'(\dfrac{(\xi - \eta)}{\sqrt{\nu}}) \rangle \begin{pmatrix} 0 & 0 \\ \sin{2\eta} & \cos{2\eta} - 1 \end{pmatrix} d\eta \bigg] \begin{pmatrix} \langle f \rangle \\ \langle g \rangle \end{pmatrix}.
\end{equation}

Assuming the Uhlenbeck-Ornstein model \eqref{eqn:autocorrelation}, 
%
(\ref{eqn:autocorrelation-b-prime}) gives
\begin{equation}\label{eqn:b-prime-centroid}
 	\langle b'(\dfrac{\xi}{\sqrt{\nu}}) b'(\dfrac{(\xi - \eta)}{\sqrt{\nu}}) \rangle = -\sigma^2\lambda^2e^{- \dfrac{\lambda\eta}{\sqrt{\nu}}}.
\end{equation}

\noindent
Using (\ref{eqn:b-prime-centroid}), eqn(\ref{eqn:matrix-2}) becomes
\begin{equation}\label{eqn:matrix-3}
	\frac{d}{d\xi} \begin{pmatrix} \langle f \rangle \\ \langle g \rangle \end{pmatrix} = \bigg[ \begin{pmatrix} 0 & 1 \\ -1 & 0 \end{pmatrix} + \frac{1}{2} \begin{pmatrix} 0 & 0 \\ c_1 & c_2 \end{pmatrix} \bigg] \begin{pmatrix} \langle f \rangle \\ \langle g \rangle \end{pmatrix},
\end{equation}
where,
\begin{equation*}
\begin{dcases}
	c_1 \equiv \frac{\sigma^2\lambda^2}{4\nu^2} \int_0^\infty e^{-\dfrac{\lambda\eta}{\sqrt{\nu}}}\sin{2\eta}d\eta \vspace{4pt}\\
	c_2 \equiv \frac{\sigma^2\lambda^2}{4\nu^2} \int_0^\infty e^{-\dfrac{\lambda\eta}{\sqrt{\nu}}}(1 - \cos{2\eta})d\eta.
\end{dcases}
\end{equation*}

\noindent
Upon integrating, we find that
\begin{equation}\label{eqn:defn-cs}
	\begin{aligned}
		c_1 &= \frac{\sigma^2\lambda^2}{2\nu(\lambda^2 + 4\nu)} \vspace{4pt}\\
		c_2 &= \frac{\sigma^2\lambda}{\sqrt{\nu}(\lambda^2 + 4\nu)}
	\end{aligned}
\end{equation}

Eqn(\ref{eqn:matrix-3}) may be rewritten as a single second order equation:
\begin{equation}\label{eqn:matrix-return-to-second-order}
	\frac{d^2\langle f\rangle}{d\xi^2} + \frac{c_2}{2}\frac{d\langle f\rangle}{d\xi} + (1 - \frac{c_1}{2})\langle f\rangle = 0,
\end{equation}
or, in terms of the original variable $t$,
\begin{equation}
	\frac{d^2\langle f\rangle}{dt^2} + \frac{c_2 \sqrt{\nu}}{2} \frac{d\langle f\rangle}{dt} + \nu (1 - \frac{c_1}{2})\langle f\rangle = 0.
\end{equation}

Putting,
\begin{equation}
	\langle f\rangle \sim e^{rt},
\end{equation}
the characteristic equation for the exponent $r$ is given by,

\begin{equation}
	r^2 + \frac{\sqrt{\nu}}{2} c_2 r + \nu(1 - \frac{c_1}{2}) = 0,
\end{equation}
from which,
\begin{equation}
	r = -\frac{\sqrt{\nu}}{4}c_2 \pm i\sqrt{\nu(1 - \frac{1}{16}c_2^2-\frac{1}{2}c_1)}.
\end{equation}

Thus, we have
\begin{equation}\label{eqn:f-simplified-coeffs}
	\langle f \rangle = e^{-\alpha t}(d_1\cos{\beta t} + d_2\sin{\beta t}).
\end{equation}
where,
$$\alpha \equiv \frac{\sqrt{\nu}}{4}c_2 \text{ and } \beta \equiv \sqrt{\nu(1 - \frac{1}{16}c_2^2-\frac{1}{2}c_1)}.$$

Substituting the expressions for $c_1$ and $c_2$ from \eqref{eqn:defn-cs}, we have
\begin{equation}
	\begin{aligned}
		&\alpha = \frac{\sigma^2\lambda}{4(\lambda^2 + 4 \nu)} \vspace{4pt}\\ \vspace{4pt}
		&\beta = \sqrt{\nu - \frac{\sigma^4\lambda^2}{16(\lambda^2 + 4\nu)^2} - \frac{\sigma^2\lambda^2}{4(\lambda^2 + 4\nu)}}
	\end{aligned}
\end{equation}

In the large-$\lambda$ limit, (\ref{eqn:f-simplified-coeffs}) becomes
\begin{equation}
	\langle f \rangle \sim e^{-\dfrac{\sigma^2}{4\lambda}t}(a_1\cos{\beta t} + a_2\sin{\beta t}), \, \beta \approx \sqrt{\nu - \dfrac{\sigma^2}{4}}.
\end{equation}

On the other hand, in the small-$\lambda$ limit, (\ref{eqn:f-simplified-coeffs}) becomes
\begin{equation}
	\langle f \rangle \sim e^{-\dfrac{\sigma^2\lambda}{16\nu}t} \left(d_1\cos{\sqrt{\nu} t} + d_2\sin{\sqrt{\nu} t} \right).
\end{equation}

\end{appendices}

\end{document}